\documentclass[12pt]{iopart}

\usepackage{iopams}
\usepackage{amsthm}
\usepackage{graphicx}
\usepackage{citesort}

\theoremstyle{definition}

\newcommand{\V}{\mathbf{V}}
\newcommand{\pr}{\mathbf{pr}}
\newcommand{\p}[1]{\partial_{#1}}

\begin{document}


\title{Discretization of partial differential equations preserving
their physical symmetries}

\author{F Valiquette\footnote{Present address: Department of Mathematics, University of
Minnesota, 526 Vincent Hall, 206 Church St. S.E., Minneapolis, MN 55455, USA.  E-mail: valiq001@math.umn.edu.} and P Winternitz}

\address{Centre de Recherches Math\'ematiques, Universit\'e de Montr\'eal,
C.P. 6128, succ. Centre-ville, Montr\'eal, QC, H3C 3J7, Canada}
\eads{\mailto{valiquet@crm.umontreal.ca} and \mailto{wintern@crm.umontreal.ca}}

\begin{abstract}
A procedure for obtaining a ``minimal'' discretization of
a partial differential equation, preserving all of its Lie point
symmetries is presented.  ``Minimal'' in this case means that the 
differential equation is replaced by a partial difference scheme
involving $N$ difference equations, where $N$ is the
number of independent and dependent variables.  We restrict
to one scalar function of two independent variables.  As examples,
invariant discretizations of the heat, Burgers and Korteweg-de
Vries equations are presented.  Some exact  solutions of the
discrete schemes are obtained.
\end{abstract}


\section{Introduction}

The dynamics of most physical processes is described by differential
equations.  Lie group theory provides powerful tools for obtaining exact
analytical solutions of such equations, specially the most fundamental
ones that are usually nonlinear (the Einstein equations, the Yang-Mills
equations, the Navier-Stokes equations, $\ldots$).

It is however quite possible that our world is actually discrete, at least at the
microscopic level.  If this is so then the fundamental equations are
finite difference ones and the differential equations obtained in
the continuous limit are actually approximations.  Even if this is not
so and no fundamental minimal space and time intervals exist, 
difference equations still play an important role in physics.  On
one hand, many physical phenomena are inherently discrete, such
as vibrations in molecular and atomic chains, or phenomena in
crystals.  On the other hand, numerical methods for solving differential
equations involve their discretization:  the differential equations are
replaced by difference ones, written on a lattice and these are then
solved.

In many cases, the symmetries of differential equations are more important
and better known than the equations themselves since these symmetries
reflect fundamental physical laws.  Thus, when discretizing, or otherwise
modifying dynamical equations, it is of interest to preserve the 
original symmetries.

This article is part of a general program that could be called ``continuous symmetries
of discrete equations'' and that has been vigorously developed for the last 
15 years, or so.  Its overall aim is to turn Lie group theory into a tool for solving
difference equations, just as it is for differential ones \cite{BDK-1997,BD-2001,D-1991,D-1994,D-2001,DK-1997,DK-2003,DKW-2000,DKW-2004,DW-2000,FNNV-1996,LTW-,LTW-2000,LTW-2004,LTW-2001,LVW-1997,LW-1991,M-1987,LW-2005,QCS-1992,RW-2004,V-2005-1,V-2005-2,W-2004}.  For
recent reviews with extensive lists of references, see \cite{LW-2005,W-2004}.

Different types of problems are treated in this program.  
A difference equation and a lattice may be given and the aim then is to 
solve the equations, using point symmetries, or generalized symmetries, as the case may be.  Alternatively, as in this article, a differential equation may 
be given and the aim is to discretize it while preserving its important qualitative features, such as point symmetries.

A formalism that unifies different approaches to symmetries of difference
systems is one in which the difference equation and the lattice are
described by a system of equations.  These relate the dependent 
and independent variables evaluated in different points.  This
has been particularly fruitful for the discretization of ordinary
differential equations (ODEs) \cite{DKW-2000,DKW-2004,RW-2004}.  

The purpose of this article is to extend this approach to the case of
partial differential equations (PDEs).  We wish to approximate them
by partial differential schemes (P$\Delta$Ss) allowing the same Lie
point symmetry group as the original PDEs.  
The schemes will be compared with related ones already
existing in the literature 
\cite{BDK-1997,BD-2001,D-1991,D-1994,D-2001,DK-1997,DK-2003,DW-2000,LTW-2001}.

A difference scheme of order
$K$ for an ODE consists of two equations relating $K+1$ points
$x_k$ and $K+1$ values $u_k$ (the discrete approximation of
the function $u(x)$ in $x_k$, evaluated in these
points):
\begin{eqnarray}
\eqalign{
E_a(\{x_{n+j},u_{n+j}\}_{j\in J})=0,\qquad a=1,2,\\
J=\{M,M+1,\ldots,N-1,N\},\qquad
n,M,N\in \mathbb{Z},\qquad N>M.}
\label{ODE scheme}
\end{eqnarray}
The points $x_k$ are distributed along a line (the $x$-axis) and the
two equations \eref{ODE scheme} must be such that if $(x_{n+M},
\ldots,x_{n+N-1},u_{n+M},\ldots,u_{n+N-1})$ are given, it is possible
to calculate $\{x_{n+N},u_{n+N}\}$.

The solution of the system will have the form
\numparts
\begin{eqnarray}
x_n&=x(n,c_1,\ldots,c_{2K}),\label{ODE lattice}\\
u_n&=u(n,c_1,\ldots,c_{2K})
\end{eqnarray}
\endnumparts
where $c_1,\ldots,c_{2K}$ are integration constants.

In the continuous limit, one of equations \eref{ODE scheme} goes
to an ODE, the other to an identity (like $0=0$).

An essential observation is that in this approach the actual lattice
\eref{ODE lattice} is not a priori given, but is obtained by solving
the system \eref{ODE scheme}.

Point transformations for the system \eref{ODE scheme} will have
exactly the same form as for the ODE, namely
\begin{equation}
\widetilde{x}=\Lambda_g(x,u),\qquad \widetilde{u}=\Omega_g(x,u),
\label{ODE group transformations}
\end{equation} 
where $g$ represents the group parameters.  They are generated by
a Lie algebra of vector fields of the form
\begin{equation}
\V=\xi(x,u)\p{x}+\phi(x,u)\p{u}.
\end{equation}
The prolongation of the vector field $\V$ to all points figuring in the 
scheme \eref{ODE scheme} has the form
\begin{equation}
\pr\;\V=\sum^{n+N}_{k=n+M}\xi(x_k,u_k)\p{x_k}+\phi(x_k,u_k)\p{u_k}
\end{equation}
and the fact that the corresponding group transformations 
\eref{ODE group transformations} will take solutions into
solutions is assured by imposing
\begin{equation}
\pr\;\V [E_a]\Big|_{E_1=E_2=0}=0,\qquad a=1,2.
\end{equation}

This approach has proven to be fruitful for large classes of ordinary
difference schemes (O$\Delta$Ss) \cite{DKW-2000,DKW-2004,LTW-2000,RW-2004}.
It provides exact discretizations of first order ODEs \cite{RW-2004}
(i.e. O$\Delta$Ss that have exactly the same solutions as their continuous
limits) and second order O$\Delta$Ss that can be exactly solved 
\cite{DKW-2000,DKW-2004}.  As pointed out earlier 
\cite{LTW-2001,LW-2005,W-2004} the use of point symmetries on fixed,
nontransforming lattices is much less fruitful.

In Section \ref{section discretization} we present the general symmetry preserving
discretization of a scalar PDE with two independent variables.  In spirit the
method is the same as used for ODEs \cite{DKW-2000,DKW-2004} and it leads to a system of 3
difference equations (rather than 2 as for ODEs).  Sections \ref{heat section},
\ref{burgers section} and \ref{kdv section} are devoted to examples.  The
linear heat equation is treated in Section \ref{heat section}, the Burgers and 
Korteweg-de Vries equations in Sections \ref{burgers section} and \ref{kdv section},
respectively.  The final Section \ref{conclusions section} is devoted to some
conclusions and the future outlook.

\section{Invariant discretization of a partial differential equation}
\label{section discretization}

\begin{figure}
\setlength{\unitlength}{1mm}
\begin{center}
\begin{picture}(130,70)
\put(0,10){\vector(1,0){130}}
\put(128,5){$x$}
\put(20,0){\vector(0,1){70}}
\put(17,68){$t$}
\qbezier(0,5)(65,25)(130,30)
\qbezier(0,25)(65,45)(130,50)
\qbezier(0,45)(65,65)(130,70)
\qbezier(15,70)(45,35)(55,0)
\qbezier(50,70)(80,35)(90,0)
\qbezier(85,70)(115,35)(125,0)
\put(32,50){$P_{m+1,n-1}$}
\put(43,32){$P_{m,n-1}$}
\put(51,14){$P_{m-1,n-1}$}
\put(62,57){$P_{m+1,n}$}
\put(74,38){$P_{m,n}$}
\put(84,20){$P_{m-1,n}$}
\put(92,62){$P_{m+1,n+1}$}
\put(106,44){$P_{m,n+1}$}
\put(116,25){$P_{m-1,n+1}$}
\put(56,0){$n-1$}
\put(91,0){$n$}
\put(126,0){$n+1$}
\put(0,1){$m-1$}
\put(0,22){$m$}
\put(0,42){$m+1$}
\end{picture}
\caption{The $(x,t)$ coordinates of points on a two-dimensional lattice.}
\label{general lattice}
\end{center}
\end{figure}
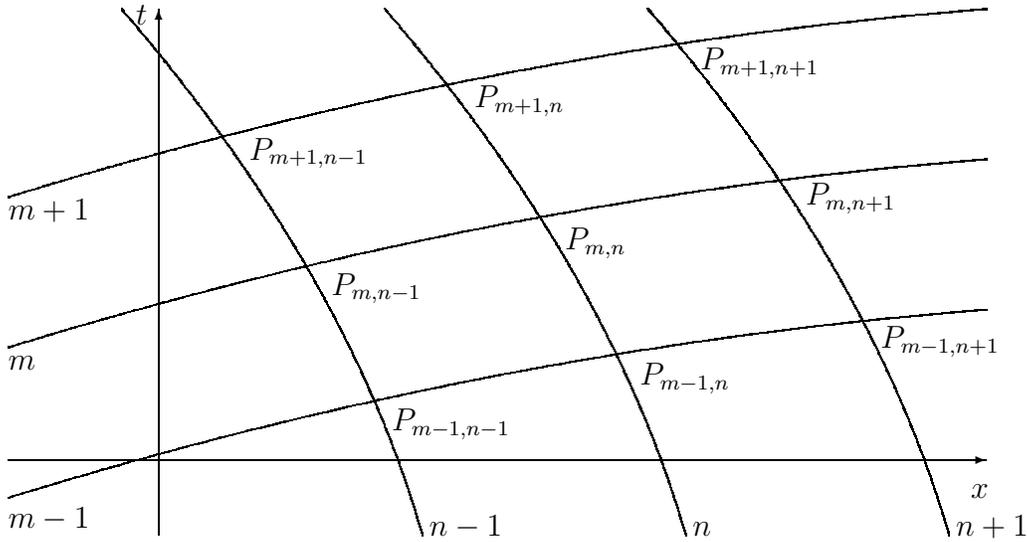

For simplicity of notation we restrict ourselves to a PDE involving
one scalar function of two variables $u(x,t)$.  It will be approximated
by a difference equation on a symmetry adapted lattice.  The lattice
consists of points distributed in a plane.  We will label these points
by an ordered pair of integers $P_{m,n}$.  We also introduce continuous
coordinates in the plane and call them $(x,t)$, though they do not necessarily
correspond to space and time and are not necessarily cartesian coordinates.
The coordinates of the point $P_{m,n}$ will be
\begin{equation}
(x_{m,n},t_{m,n}),\qquad (m,n)\in\mathbb{Z}^2,
\end{equation}
see figure \ref{general lattice}.

The actual partial difference scheme will be a set of relations between the 
variables $(x_{m,n},t_{m,n},u_{m,n})$
evaluated at a finite number of points on the lattice.  The first
question is:  how many relations and how many points do we 
need?

We start from a given PDE
\begin{equation}
E(x,t,u^{(n)}(x,t))=0,\label{pde}
\end{equation}
where $u^{(n)}(x,t)$ denotes all the partial derivatives of $u(x.t)$ up
to order $n$.  We assume that equation \eref{pde} is invariant
under a group $G$ of local Lie point transformations with a Lie algebra
$L$ realized by vector fields of the form 
\begin{equation}
\V=\xi(x,t,u)\p{x}+\eta(x,t,u)\p{t}+\phi(x,t,u)\p{u}.\label{vector field}
\end{equation} 
We wish to approximate the PDE \eref{pde} by a system
of finite difference equations
\begin{eqnarray}
\eqalign{
E_a(\{x_{m+j_1,n+j_2},t_{m+j_1,n+j_2},u_{m+j_1,n+j_2}\}_{(j_1,j_2)\in
  J})=0,\\
1\leq a \leq N,\qquad {\rm where}\;\{0\}\subset J
\subset \mathbb{Z}^2,}
\label{pfde}
\end{eqnarray}  
relating the quantities $(x,t,u)$ at a finite number of points and invariant
under the same group $G$ as  \eref{pde}.  The minimal number
of equations needed in \eref{pfde} is $N=3$, determining the values
of $u$, $x$ and $t$ at different points.  If only 3 equations are imposed,
then the solution of  \eref{pfde} will depend on a certain number of 
arbitrary functions of one variable ($m$, $n$ or some combination
of $m$ and $n$).  How many such functions will be present depends on
the order of the system, i.e. the number of points corresponding to
the set $J$.  This in turn
depends on the order of the PDE \eref{pde} and the precision we are looking for.

It may be convenient to impose more than 3 equations, to specify
the lattice to a larger degree.  For example, in reference \cite{LTW-2001}
the number of equations imposed was 5.  Four of them specified
the mesh and made it possible to move along the lattice in all directions
from any point.  The additional equations $(N>3)$ cause the system
\eref{pfde} to be overdetermined, so some compatibility conditions
must be
satisfied.  These additional equations, if imposed, will play the same
role as initial conditions, or boundary conditions for PDEs:  they will
partially, or completely specify the arbitrary functions involved.

To illustrate the point, consider the linear PDE
\begin{equation}
u_{xt}=0.\label{example pde}
\end{equation}
We approximate it by the P$\Delta$S \cite{LW-2005}
\begin{eqnarray}
\label{example scheme}
\eqalign{
E_1&=\frac{1}{t_{m+1,n}-t_{m,n}}\left(
\frac{u_{m+1,n+1} - u_{m+1,n}}{x_{m+1,n+1}-x_{m+1,n}} -
\frac{u_{m,n+1} - u_{m,n}}{x_{m,n+1}-x_{m,n}}
\right)=0,\\
E_2&=t_{m,n+1}-t_{m,n}=0\\
E_3&=x_{m+1,n}-x_{m,n}=0.}
\end{eqnarray}
This system can be solved explicitly to obtain
\numparts
\begin{eqnarray}
t_{m,n}=\alpha(m),\qquad x_{m,n}=\beta(n),\label{example lattice}\\
u_{m,n}=f(x_{m,n})+g(t_{m,n}).\label{example solution}
\end{eqnarray}
\endnumparts
Thus the general solution involves 4 arbitrary functions of 1 variable
each.  The lattice is orthogonal, the spacing of points in the two 
directions is arbitrary.  We mention that the P$\Delta$S \eref{example scheme} is invariant under an arbitrary reparametrization of $t$ and
$x$ (i.e. of $m$ and $n$), corresponding to the invariance
of the PDE  \eref{example pde} under the infinite-dimensional group of
conformal transformations.  If we wish to impose that the lattice
be uniform in both directions, we can add two further equations
\begin{eqnarray}
\label{example restriction}
\eqalign{
E_4&=t_{m+1,n}-2t_{m,n}+t_{m-1,n}=0,\\
E_5&=x_{m,n+1}-2x_{m,n}+x_{m,n-1}=0.}
\end{eqnarray}
The solution \eref{example lattice} then restricts to
\begin{equation}
t_{m,n}=\tau\;m+t_0,\qquad x_{m,n}=h\;n+x_0,
\end{equation}
with $u_{m,n}$ as in \eref{example solution} ($t_0$, $\tau$,
$x_0$, $h$ are constants).  The P$\Delta$S given by
\eref{example scheme} and \eref{example restriction} is
no longer invariant under the infinite-dimensional conformal group.
The symmetry group of the lattice is reduced to 
dilations and translations of $t$ and $x$.

Since we wish to obtain a symmetry preserving discretization of
the PDE, the P$\Delta$S \eref{pfde} must
be constructed out of invariants and invariant manifolds of the symmetry
group $G$.  The procedure is standard \cite{BDK-1997,BD-2001,D-1991,D-1994,D-2001,DK-1997,DK-2003,DKW-2000,DKW-2004,DW-2000,LW-2005,W-2004}
and we describe it only briefly.

\vskip 0.25cm
\noindent 1.\hskip 0.5cm 
We specify a stencil i.e. choose the number and positions of the points
to be used in the P$\Delta$S.

\vskip 0.25cm
\noindent 2.\hskip 0.5cm
We prolong the vector field \eref{vector field} to all points of the stencil
(i.e. all the points figuring in the P$\Delta$S)
\begin{equation}
\fl \pr\;\V=\sum_{J}\left\{ \xi_{m+j_1,n+j_2}\p{x_{m+j_1,n+j_2}}+
 \eta_{m+j_1,n+j_2}\p{t_{m+j_1,n+j_2}}
 + \phi_{m+j_1,n+j_2}\p{u_{m+j_1,n+j_2}} \right\},
\label{discrete prolongation}
\end{equation}
where $\xi_{m+j_1,n+j_2}=\xi(x_{m+j_1,n+j_2},t_{m+j_1,n+j_2},u_{m+j_1,n+j_2})$ and similarly for $\eta$ and $\phi$.

\vskip 0.25cm
\noindent 3. \hskip 0.5cm
We find the elementary invariants of $G$ by solving the system of first
order PDEs
\begin{equation}
\pr\; \V[ I(\{x_{m+j_1,n+j_2},t_{m+j_1,n+j_2},u_{m+j_1,n+j_2}\}_{(j_1,j_2)\in
J})]=0,\label{infinitesimal invariance}
\end{equation}
where $\V$ is a general element of the symmetry algebra $L$ of the
PDE \eref{pde} (but the prolongation is the ``discrete'' one 
\eref{discrete prolongation}).  The algebra may be finite, or 
infinite dimensional.  If we have dim $L=l<\infty$ then we choose
a convenient basis $\V_a$, $a=1,2,\ldots,l$ and \eref{infinitesimal invariance} reduces to a system of $l$ linear first order PDEs.)
Using the method of characteristics, 
we obtain a set of elementary invariants
$I_1,\ldots,I_\mu$.  Their number is given by the formula
\begin{equation}
\mu={\rm dim}\; M - {\rm rank}\; Z,
\end{equation}
where $M$ is the manifold that $G$ acts on, i.e. 
\begin{equation}
M\sim\{\{x_{m+j_1,n+j_2},t_{m+j_1,n+j_2},u_{m+j_1,n+j_2}\}_{(j_1,j_2)\in
J}\}.
\end{equation}
Thus ${\rm dim}\;M= N \times \#J$,  where $\#J$ denotes 
the order of the set $J$ and $Z$ is the matrix
\begin{equation}
Z=
\left(\begin{array}{c}
\{\xi^1_{m+j_1,n+j_2},\eta^1_{m+j_1,n+j_2},\phi^1_{m+j_1,n+j_2}\}_{(j_1,j_2)\in J}\\
\vdots\\
\{\xi^l_{m+j_1,n+j_2},\eta^l_{m+j_1,n+j_2},\phi^l_{m+j_1,n+j_2}\}_{(j_1,j_2)\in J}
\end{array} \right)
\label{Z matrix}
\end{equation}
formed with the coefficients 
of the prolonged symmetry generators $\V_a$
generating the basis of the finite dimensional Lie algebra. 
  
Since the quantities $I_1,\ldots,I_\mu$ form a basis
of elementary invariants, any
difference equation 
\begin{equation}
E(I_1,\ldots,I_\mu)=0\label{strong invariant}
\end{equation}
will be invariant under the group $G$.
The equation \eref{strong invariant} obtained in this 
manner is said to
be strongly invariant and satisfies $\pr\; \V_a[E]=0,$ $a=1,\ldots,l$, 
identically.

Other invariant equations can be obtained if the rank of the matrix 
$Z$ is not maximal on some manifolds described by equations of the
form $E(\{x_{m+j_1,n+j_2},t_{m+j_1,n+j_2},u_{m+j_1,n+j_2}
\}_{(j_1,j_2)\in J})=0$ that
satisfy
\begin{equation}
\pr\; \V_a[E]\Big|_{E=0}=0,\qquad a=1,\ldots,l.
\label{weakly invariance} 
\end{equation}
Such equations are said to be weakly 
invariant.  In practice we can start by computing the 
invariant manifolds and this can facilitate the computation of the 
set of fundamental invariants.

In this article we will always choose the minimal number of equations
needed, i.e. $N=3$ in \eref{pfde}.  By construction these equations
all satisfy
\begin{equation}
\pr\;\V[E_a]\Big|_{E_1=E_2=E_3=0}=0,\qquad a=1,2,3,
\end{equation}
i.e. they are invariant under the group $G$.

When constructing the P$\Delta$S out of the invariants
$I_1,\ldots,I_\mu$ and the invariant manifolds some choices
must be made.  The most important constraint is given by the
continuous limit.  We impose that $E_1=0$ reduces
to the PDE \eref{pde} and $E_2=E_3=0$ reduce to identities.
Further constraints may come from the boundaries, or initial
conditions imposed on solutions of the PDE that we are solving,
or from the precision that we desire.

The calculation of group invariants described above is completely
analogous to the one used in the continuous case \cite{O-1993,O-1995}.
The Lie algebra approach that we use can be replaced by the Lie group one using moving frames \cite{FO-1998,O-2003}.

If the symmetry group of the original PDE is infinite-dimensional, some
modifications of the procedure are required.  In particular, if the PDE is
linear, then an infinite-dimensional pseudogroup corresponding to
the linear superposition principle is always present.  In this case we can
restrict ourselves to invariants $I_k$ of the finite dimensional
subgroup of the symmetry group and then require that the P$\Delta$S
formed out of the invariants be linear in $u$.

\section{The linear heat equations}\label{heat section}

The linear heat equation
\begin{equation}
u_t=u_{xx}\label{heat equation}
\end{equation}
is a much used example when Lie group theory is applied
to the study of differential equations.  Its symmetry group was
already known to Sophus Lie and is reproduced in virtually every book
on the subject (see e.g. \cite{O-1993}).  A basis for its symmetry algebra
consists of
\numparts
\label{heat algebra}
\begin{eqnarray}
\eqalign{
\V_1  =  \partial_x,\qquad
\V_2  =  \partial_t,\qquad 
\V_3  =  u\partial_u,\qquad
\V_4  =  x\partial_x + 2t\partial_t, \\
\V_5  =  2t\partial_x - xu\partial_u,\qquad
\V_6  =  4tx\partial_x + 4t^2\partial_t - (x^2+2t)u\partial_u,}
\label{heat finite algebra}
\end{eqnarray}
\begin{equation}
\V_\alpha = \alpha(x,t)\partial_u \qquad\rm{where}\qquad
            \alpha_t=\alpha_{xx}.\label{heat infinite algebra}
\end{equation}
\endnumparts
Like everybody else, we use the heat equation \eref{heat equation}
as an example, precisely because it is linear, as reflected in the infinite
dimensional algebra \eref{heat infinite algebra} and has a large
and interesting finite dimensional subalgebra \eref{heat finite algebra}
of the symmetry algebra $L$.

Our aim is to discretize  \eref{heat equation} while preserving the 
entire Lie point symmetry algebra (25).  
Several similar discretizations exist in the 
literature.  In \cite{BDK-1997} and \cite{DK-2003} the authors
first introduce Lagrangian variables and represent the equation
\eref{heat equation} as a system of two equations.  These are then
discretized preserving all the symmetries \eref{heat finite algebra},
not however linearity.  A different point of view was taken in \cite{FNNV-1996,LVW-1997,LTW-2004}.  Equation \eref{heat equation}
(and any other linear equation) is discretized on a uniform orthogonal
nontransforming lattice.  Linearity is preserved, but the Lie
algebra of point symmetries (25) is replaced
by an isomorphic Lie algebra of difference operators.  Thus,
the Lie point character of the transformations is given up while all the
algebraic consequences of Lie symmetry are preserved. 
Here we shall directly apply the procedure outlined in Section \ref{section discretization}.

\subsection{Invariant schemes for the linear heat equation}

Before computing a set of elementary discrete invariants, let us
introduce a convenient notation for points on the lattice, to be used
for the heat equation and all further examples in this article.  We put:
\begin{eqnarray}
\eqalign{
(x_{m,n},t_{m,n},u_{m,n})\equiv(x,t,u),\\
(x_{m+1,n},t_{m+1,n},u_{m+1,n})\equiv(\hat{x},\hat{t},\hat{u}),\\
(x_{m,n\pm 1},t_{m,n\pm 1},u_{m,n\pm 1})\equiv(x_\pm,t_\pm,u_\pm),\\
(x_{m,n\pm 2},t_{m,n\pm 2},u_{m,n\pm 2})\equiv(x_{\pm\pm}
,t_{\pm\pm},u_{\pm\pm}),\\
(x_{m+1,n\pm 1},t_{m+1,n\pm 1},u_{m+1,n\pm 1})
\equiv(\hat{x}_\pm,\hat{t}_\pm,\hat{u}_\pm),\\
(x_{m+1,n\pm 2},t_{m+1,n\pm 2},u_{m+1,n\pm 2})\equiv(\hat{x}_{\pm\pm}
,\hat{t}_{\pm\pm},\hat{u}_{\pm\pm}),}
\label{notation}
\end{eqnarray}
and introduce the steps 
\begin{eqnarray}
\eqalign{
h_\pm=\pm(x_\pm-x),   \qquad h_{\pm\pm}=\pm(x_{\pm\pm}-x_\pm),  \qquad
\hat{h}_\pm=\pm(\hat{x}_\pm-\hat{x}), \\
\hat{h}_{\pm\pm}=\pm(\hat{x}_{\pm\pm}-\hat{x}_\pm), \qquad
\sigma=\hat{x}-x, \qquad  \sigma_+=\hat{x}_+-x_+, \\
\tau=\hat{t}-t,  \qquad  T_\pm=\pm(t_\pm-t).} 
\end{eqnarray}
(this notation is standard in numerical analysis).

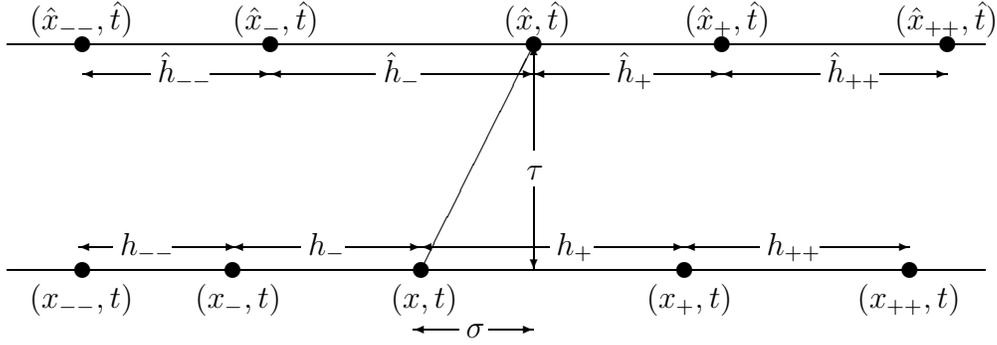
\begin{figure}
\setlength{\unitlength}{1mm}
\begin{center}
\begin{picture}(130,50)
\put(0,10){\line(1,0){130}}
\put(0,40){\line(1,0){130}}
\put(55,10){\line(1,2){15}}
\put(10,10){\circle*{2}} \put(30,10){\circle*{2}}
\put(55,10){\circle*{2}}\put(90,10){\circle*{2}}
\put(120,10){\circle*{2}}
\put(10,40){\circle*{2}} \put(35,40){\circle*{2}}
\put(70,40){\circle*{2}}\put(95,40){\circle*{2}}
\put(125,40){\circle*{2}}
\put(3,42){($\hat{x}_{--},\hat{t}$)}
\put(30,42){($\hat{x}_-,\hat{t}$)}
\put(66,42){($\hat{x},\hat{t}$)}
\put(90,42){($\hat{x}_+,\hat{t}$)}
\put(118,42){($\hat{x}_{++},\hat{t}$)}
\put(3,5){($x_{--},t$)}
\put(25,5){($x_-,t$)} 
\put(51,5){($x,t$)}
\put(85,5){($x_+,t$)}
\put(113,5){($x_{++},t$)}
\put(70,26){\vector(0,1){14}}
\put(69,22){$\tau$}
\put(70,21){\vector(0,-1){11}}
\put(116,36){\vector(1,0){9}}
\put(109,35){$\hat{h}_{++}$}
\put(108,36){\vector(-1,0){13}}
\put(86,36){\vector(1,0){9}}
\put(81,35){$\hat{h}_+$}
\put(80,36){\vector(-1,0){10}}
\put(55,36){\vector(1,0){15}}
\put(50,35){$\hat{h}_-$}
\put(49,36){\vector(-1,0){14}}
\put(27,36){\vector(1,0){8}}
\put(20,35){$\hat{h}_{--}$}
\put(19,36){\vector(-1,0){9}}
\put(22,13){\vector(1,0){8.5}}
\put(15,12){$h_{--}$}
\put(14,13){\vector(-1,0){4.5}}
\put(45,13){\vector(1,0){10}}
\put(40,12){$h_-$}
\put(39,13){\vector(-1,0){9}}
\put(78,13){\vector(1,0){12}}
\put(73,12){$h_+$}
\put(72,13){\vector(-1,0){17}}
\put(108,13){\vector(1,0){12}}
\put(101,12){$h_{++}$}
\put(100,13){\vector(-1,0){10}}
\put(64,2){\vector(1,0){6}}
\put(61,1){$\sigma$}
\put(60,2){\vector(-1,0){6}}
\end{picture}
\caption{Lattice for the invariant discretization of
the heat and Burgers equations (6 points) and the
KdV equation (10 points).}
\label{figure lattice}
\end{center}
\end{figure}

It is easy to see that the equation
\begin{equation}
T_+=0\label{heat invariant manifold}
\end{equation}
 is invariant under the entire group $G$ generated by the
 algebra $L$ of  (25)
 (i.e. we have $\pr\;\V[T_+]|_{T_+=0}=0$ for all $\V \in L$).  We will
 include  \eref{heat invariant manifold} in all of our invariant schemes
 which means that we will always have horizontal time layers.  Equation
 \eref{heat invariant manifold} implies that in \eref{notation} we
 have
 \begin{displaymath}
 t_{++}=t_+=t=t_-=t_{--},\qquad
  \hat{t}_{++}=\hat{t}_+=\hat{t}=\hat{t}_-=\hat{t}_{--},
 \end{displaymath}
 as indicated on figure \ref{figure lattice}.   The fact that the time layers are horizontal is particularly convenient in numerical simulations.

For the heat equation we will not need all 10 points of figure
\ref{figure lattice}.  We will use 6 of them and in view of 
\eref{heat invariant manifold} we can restrict to the 14 dimensional
space with coordinates 
\begin{equation}
(x_-, x, x_+, \hat{x}_-, \hat{x},
\hat{x}_+, t, \hat{t}, u_-, u, u_+, \hat{u}_-, \hat{u}, \hat{u}_+).
\label{14 dimensional space} 
\end{equation}
Imposing invariance under the group generated by the 6
dimensional algebra \eref{heat finite algebra} we obtain 8 elementary
invariants 
\begin{eqnarray}
\label{heat invariants}
\eqalign{
I_1&=\frac{h_+}{h_-},\qquad
I_2=\frac{\hat{h}_+}{\hat{h}_-},\qquad
I_3=\frac{h_+\hat{h}_+}{\tau},\qquad
I_4=\frac{\sqrt{\tau}}{h_+}\frac{\hat{u}}{u}
\exp\bigg[\frac{\sigma^2}{4\tau}\bigg],\\
I_5&=\frac{u_+}{u}\exp\bigg[ \frac{h_+}{4\tau}(2\sigma-h_+)\bigg],
\qquad
I_6=\frac{u_-}{u}\exp\bigg[-\frac{h_-}{4\tau}(2\sigma+h_-)\bigg],\\
I_7&=\frac{\hat{u}_+}{\hat{u}}\exp\bigg[ \frac{\hat{h}_+}{4\tau}
(2\sigma+\hat{h}_+)\bigg],\qquad
I_8=\frac{\hat{u}_-}{\hat{u}}\exp\bigg[-\frac{\hat{h}_-}{4\tau}
(2\sigma-\hat{h}_+)\bigg].}
\end{eqnarray} 
This set is equivalent to the one used in references
\cite{BDK-1997,DK-2003}, but we find the set \eref{heat invariants} more convenient
when imposing linearity.

Using the invariants \eref{heat invariants} we obtain an explicit
invariant scheme that is linear in $u$ by setting
\begin{equation}
I_3^{3/2}I_4-I_3=(I_5+I_6)\exp\left[\frac{I_3}{4}\right]-2,\qquad
T_+=0,\qquad 
I_1=1.
\label{invariant explicit heat}
\end{equation}
In terms of the variables $(x,t,u)$ we have
\numparts
\begin{eqnarray}
\eqalign{
\fl\frac{1}{\tau}\bigg(\sqrt{\frac{\hat{h}}{h}}\hat{u}
\exp\bigg[\frac{\sigma}{4\tau}\bigg]-u\bigg)=\\
\frac{1}{\hat{h}h}\bigg(
u_+\exp\bigg[ \frac{h}{4\tau}(2\sigma -h+\hat{h})\bigg]-2u+
u_-\exp\bigg[-\frac{h}{4\tau}(2\sigma +h-\hat{h})\bigg]\bigg),}
\label{explicit heat u}\\
T_+=0,\label{explicit heat t}\\
h_+=h_-\equiv h.\label{explicit heat x}
\end{eqnarray}
\endnumparts

An invariant implicit scheme, also linear in $u$, is obtained by setting
\begin{equation}
I_3-I_3^{1/2}I_4^{-1}=(I_7+I_8)\exp\left[-\frac{I_3}{4}\right]-2,\qquad
T_+=0,\qquad 
I_2=1,
\label{invariant implicit heat}
\end{equation}
which in terms of the discrete variables gives
\begin{eqnarray}
\eqalign{
\fl\frac{1}{\tau}\bigg(\hat{u}-\sqrt{\frac{h}{\hat{h}}}u
\exp\bigg[-\frac{\sigma}{4\tau}\bigg]\bigg)=\\
\frac{1}{\hat{h}h}\bigg(
\hat{u}_+\exp\bigg[ \frac{\hat{h}}{4\tau}(2\sigma-h+\hat{h})
\bigg]-2\hat{u}+
\hat{u}_-\exp\bigg[-\frac{h}{4\tau}(2\sigma +h-\hat{h})
\bigg]\bigg),\\
T_+=0,\label{implicit heat}\\
\hat{h}_+=\hat{h}_-\equiv \hat{h}.}
\end{eqnarray}

We recall that in this context ``explicit'' means that
one value of $\hat{u}$ on a higher time level is express
in terms of values of $u$ at a previous time $t$.  ``Implicit''
means that several values of $\hat{u}$ at time $\hat{t}$ are calculated
simultaneously (for $\hat{x}_+$, $\hat{x}$ and $\hat{x}_-$).  

Notice that unlike for the standard explicit and implicit
discretization
\begin{equation}
\frac{\hat{u}-u}{\tau}=\frac{u_+-2u+u_-}{h^2},\qquad
\frac{\hat{u}-u}{\tau}=\frac{\hat{u}_+-2\hat{u}+\hat{u}_-}{\hat{h}^2},
\label{standard discretization heat}
\end{equation}
it is not possible in the
invariant case to have schemes involving just $h$ or $\hat{h}$.
Indeed, in the invariant schemes both $h$ and $\hat{h}$ are
present.  This is a consequence of the use of $I_3$ when
generating the schemes.  In the particular case when $\sigma$ is 
chosen to be zero, we get $h=\hat{h}$ and then
the invariant schemes reduce to the standard discretizations
\eref{standard discretization heat}
on an orthogonal lattice.  
It is important to realise that the choice $\sigma=0$ is
not an invariant one.  In fact, it is not invariant under
the transformations generated by $\V_5$ and $\V_6$.

The two equations for the lattice of the
invariant schemes (32) and
\eref{implicit heat} are easily solved and give
\begin{equation}
t_{m,n}=\gamma(m),\qquad x_{m,n}=h(m)n+x_0(m),
\label{lattice}
\end{equation}
where $\gamma(m)$, $h(m)$ and $x_0(m)$ are arbitrary functions.

\subsection{Continous limit of the invariant schemes}

We now compute the continuous limit of the explicit
invariant scheme (32) to first of 
all obtain an important condition
on the limit of the ratio $\sigma/\tau$ and show that
the scheme then coverges to the heat equation.  We will not
present the calculations for the implicit schemes since they are
quite similar.

Clearly, the equations for the lattice \eref{explicit heat t} 
and \eref{explicit heat x} go to 0=0 in the continuous limit.
We must show that \eref{explicit heat u} goes to 
\eref{heat equation}.  To compute this limit, we introduce infinitesimal
parameters that go to zero in the continous limit.  Since the
steps induced by the discrete variable $n$ must go to zero independently
of those induced by the discrete variable $m$, we must have
\begin{equation}
\gamma(m)=\epsilon\widetilde{\gamma}(m),\qquad h(m)=\delta\widetilde{h}(m),
\label{parameter tau and h}
\end{equation}
where $\epsilon$ and $\delta$ are independent infinitesimal parameters and
$\widetilde{\gamma}(m)$ and $\widetilde{h}(m)$ are finite quantities.
By the definition of $x_{m,n}$ in \eref{lattice} we have
\begin{eqnarray*} 
\sigma&=(h(m+1)-h(m))n+x_0(m+1)-x_0(m)\\
&=\delta(\widetilde{h}(m+1)
-\widetilde{h}(m))n+x_0(m+1)-x_0(m).  
\end{eqnarray*}
Since, $\sigma$ must go to zero when $\tau$ goes to zero, $\sigma$
must converge to zero as
\begin{equation}
\sigma=\delta \epsilon^k\sigma_1(m)n+\epsilon^l\sigma_2(m),  
\label{parameter sigma}
\end{equation} 
with $k,l>0$.  With these considerations we have 
\begin{displaymath}
\frac{\sigma_+-\sigma}{h} \to 0,
\end{displaymath}
in the continous limit.  By writing
\begin{equation}
\hat{h}=h+\sigma_+-\sigma,
\end{equation}  
the expansion of 
\begin{displaymath}
\sqrt{\frac {\hat{h}}{h}}=\sqrt{1+\frac{\sigma_+-\sigma}{h}},
\end{displaymath}
in a Taylor series is well defined.  By developing the left hand side of
\eref{explicit heat u} we find
\begin{equation}
\frac{1}{2}\frac{\sigma_+-\sigma}{\tau h}+\frac{\sigma^2}{4\tau^2}u
+\frac{\sigma}{\tau}u_x+u_t+\Or(\frac{1}{\tau}\left(\frac{\sigma_+
-\sigma}{h}\right)^2,\frac{\sigma^2}{\tau},\tau,\sigma).
\label{left hand side} 
\end{equation}
Equation \eref{left hand side} written in terms of the parameters
$\epsilon$ and $\delta$ gives
\begin{eqnarray}
\eqalign{
\fl\frac{1}{2}\epsilon^{k-1}
+\frac{1}{4}(\delta\epsilon^{k-1}\sigma_1(m)n+
\epsilon^{l-1}\sigma_2(m))^2u\\
+(\delta\epsilon^{k-1}\sigma_1(m)n+\epsilon^{l-1}\sigma_2(m))u_x
+u_t+\Or(\epsilon^{2k-1},\epsilon^{2l-1})}
\label{left hand side 2}
\end{eqnarray}
For the coefficient in front of $u_x$ in \eref{left hand side 2}
not to diverge in the limit we must impose
\begin{equation}
k,l\geq 1
\end{equation}
For the invariant scheme to converge we must have
\begin{equation}
\lim_{\tau\to 0}\frac{\sigma}{\tau}<\infty.\label{limit condition}
\end{equation} 

The expansion of the right hand side of \eref{explicit heat u}
in a Taylor series gives
\begin{equation}
\frac{1}{2}\frac{\sigma_+-\sigma}{\tau h}+\frac{\sigma^2}{4\tau^2}u+
\frac{\sigma}{\tau}u_x +u_{xx}+\Or(\epsilon^2k-1,\delta).
\label{right hand side}
\end{equation}
Equating \eref{left hand side} to \eref{right hand side} we get
\begin{displaymath}
u_t=u_{xx}+\Or(\epsilon^{2k-1},\epsilon^{2l-1},\delta),
\end{displaymath}
which converges to the heat equation when $\epsilon$ and
$\delta$ go to zero.

\subsection{Exact solutions of the invariant schemes}
\label{exact solution heat}

One of the interesting aspects of generating symmetry-preserving
schemes is that group theory can be used to obtain
nontrivial  exact solutions, using the 
fundamental concept that the symmetry group
maps solutions to solutions.    

One obvious solution of the schemes (32) and \eref{implicit heat}
is the linear solution 
\numparts
\begin{equation}
u_{m,n}=ax_{m,n}+b,\label{linear solution heat}
\end{equation}
defined on the orthogonal lattice
\begin{equation}
t_{m,n}=\gamma(m),\qquad x_{m,n}=hn+x_0,\label{linear lattice heat}
\end{equation}
\endnumparts
where $h$ and $x_0$ are constants.  Indeed, as mentioned
earlier, on such a lattice $\sigma=0$ so the invariant discretizations
go to the standard discretizations \eref{standard discretization heat} and
it is well known that \eref{linear solution heat} is an exact solution.
If we want, we can fix $\gamma(m)$ to be equal to $\tau m+t_0$,
where $\tau$ and $t_0$ are constants, so that the lattice is rectangular.

Given a discrete solution $(x,t,u(x,t))$ new solutions
can be obtained by acting on the known solution with the symmetry
group generated by \eref{heat finite algebra}.  Hence,
\begin{eqnarray}
\label{heat general transformation}
\fl \eqalign{
\widetilde{x}=\frac{e^{\epsilon_4}(x+\epsilon_1)+2\epsilon_5
e^{2\epsilon_4}(t+\epsilon_2)}{1-4\epsilon_6 e^{2\epsilon_4}(t+\epsilon_2)},\qquad\qquad
\widetilde{t}=\frac{e^{2\epsilon_4}(t+\epsilon_2)}
{1-4\epsilon_6 e^{2\epsilon_4}(t+\epsilon_2)},\\
\widetilde{u}(\widetilde{x},\widetilde{t})=
\frac{1}{\sqrt{1+4\epsilon_6\widetilde{t}}}\;
u\left(e^{-\epsilon_5}\frac{\widetilde{x}-2\epsilon_6\widetilde{t}}
{1+4\epsilon_6\widetilde{t}},
\frac{e^{-2\epsilon_4}\widetilde{t}}{1+4\epsilon_6\widetilde{t}}
-\epsilon_2 \right)
\exp\left[\epsilon_3-\frac{\epsilon_5 \widetilde{x}
-\epsilon_5^2 \widetilde{t}+\epsilon_6
\widetilde{x}^2}{1+4\epsilon_6 \widetilde{t}}\right],}
\end{eqnarray}
is also a solution of the invariant schemes (32) and
\eref{implicit heat}.  If the general transformation is applied to the
linear solution, we conclude that
\numparts
\begin{eqnarray}
\widetilde{x}&=\frac{e^{\epsilon_4}(hn+x_0+\epsilon_1)+2\epsilon_5
e^{2\epsilon_4}\gamma(m)}{1-4\epsilon_6 e^{2\epsilon_4}(\gamma(m)+
\epsilon_2)},\label{symmetry transformation x}\\
\widetilde{t}&=\frac{e^{2\epsilon_4}(\gamma(m)+\epsilon_2)}
{1-4\epsilon_6 e^{2\epsilon_4}(\gamma(m)+\epsilon_2)},\label{symmetry transformation t}\\
\widetilde{u}(\widetilde{x},\widetilde{t})&=
\frac{1}{\sqrt{1+4\epsilon_6\widetilde{t}}}
(ae^{-\epsilon_5}\frac{\widetilde{x}-2\epsilon_6\widetilde{t}}
{1+4\epsilon_6\widetilde{t}}+b)
\exp\left[\epsilon_3-\frac{\epsilon_5 \widetilde{x}
-\epsilon_5^2 \widetilde{t}+\epsilon_6
\widetilde{x}^2}{1+4\epsilon_6 \widetilde{t}}\right],
\end{eqnarray}
\endnumparts
is a solution.  

Furthermore, the limit condition \eref{limit condition} for the
ratio $\sigma/\tau$ is preserved under the general 
symmetry transformation \eref{symmetry transformation x} and
\eref{symmetry transformation t},
since  
\begin{equation}
\frac{\widetilde{\sigma}}{\widetilde{\tau}}=
\frac{\hat{\widetilde{x}}-\widetilde{x}}{\hat{\widetilde{t}}-\widetilde{t}}=
\frac{\sigma}{\tau}(e^{-\epsilon_4}-4\epsilon_6e^{\epsilon_4}(t+\epsilon_2))
+2\epsilon_5+4\epsilon_6e^{\epsilon_4}(x+\epsilon_1)
\end{equation}
which does not diverge in the continuous limit by 
hypothesis on the ratio $\sigma/\tau$.

\begin{figure}[t]
\begin{center}
\includegraphics[height=8cm,width=10cm]{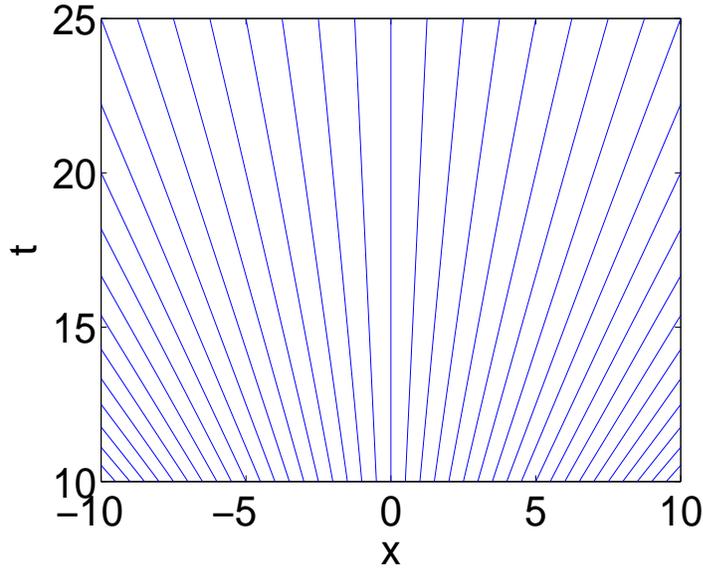}
\end{center}
\caption{Lattice for the fundamental solution of the heat equation, 
$h=0.05$, $x_0=0$, $\tau=0.005$, $t_0=10$.}
\label{figure lattice fondamental solution}
\end{figure}

The fundamental solution of the heat equation 
can be obtained by successively applying the symmetry transformations
$\exp[\epsilon\V_6]$, $\exp[\ln(\sqrt{\epsilon/\pi b^2})\V_3]$ and
$\exp[1/(4\epsilon)\V_2]$ to the constant solution $u=b\neq 0$
and on the lattice $x=hn+x_0$, $t=\gamma(m)$ we get
\numparts
\begin{equation}
\widetilde{u}=\sqrt{\frac{1}{4\pi \widetilde{t}}}
\exp\left[\frac{-\widetilde{x}^2}{4\widetilde{t}}\right]
\end{equation}
defined on the lattice
\begin{equation}
\widetilde{x}=\frac{x}{1-4\epsilon t},\qquad \widetilde{t}
=\frac{1}{4\epsilon(1-4\epsilon t)}.
\end{equation}
\endnumparts
By taking $\epsilon=1/4$ and $\gamma(m)=\frac{1}{4}
\left(1-\frac{1}{m\tau+t_0}\right)$, we get the lattice
\begin{equation}
\widetilde{x}=(hn+x_0)(\tau m+t_0),\qquad \widetilde{t}=\tau m+t_0,
\label{maillage solution fondamentale}
\end{equation}
where $h$, $x_0$, $\tau$ and $t_0$ are constants.
The evolution of the lattice is shown in figure \ref{figure lattice fondamental solution}.

Another interesting solution that can be obtained is the 
traveling wave solution.  This one is obtained by applying the
symmetry transformation $\exp[c\V_5]$
\numparts
\begin{equation}
\widetilde{u}=(a(\widetilde{x}-2c\widetilde{t})+b)
\exp[-c\widetilde{x}+c^2\widetilde{t}],
\end{equation}
and is defined on the lattice
\begin{equation}
\widetilde{x}=hn+x_0+2c\widetilde{t},\qquad
\widetilde{t}=\gamma(m).
\end{equation}
\endnumparts
By taking $\gamma(m)=\tau m +t_0$, we obtain a lattice
in which the points in $x$ move linearly as a function of
time.

\section{Burgers' equation}\label{burgers section}

In this section we shall derive invariant schemes for
the Burgers' equation
\begin{equation}
v_t+vv_x=v_{xx}\label{burgers equation}.
\end{equation}
This equation is well known to be related to the heat
equation \eref{heat equation} by the Cole-Hopf 
transformation
\begin{equation}
v(x,t)=-2\frac{u_x(x,t)}{u(x,t)}.\label{cole-hopf}
\end{equation}
Since the transformation \eref{cole-hopf} is not a point one,
the symmetry algebras of the Burgers' equation
and the heat equation are not isomorphic.  Indeed, the
symmetry algebra of \eref{burgers equation} is five dimensional 
and is spanned by the vector fields \cite{O-1993}
\begin{eqnarray}
\label{burgers algebra}
\eqalign{
\V_1=\partial_x,\qquad
\V_2=\partial_t,\qquad
\V_3=t \partial_x + \partial_v,\qquad\qquad\\
\V_4=x \partial_x + 2t \partial_t - v \partial_v,\qquad
\V_5=tx \partial_x + t^2 \partial_t + (x-tv) \partial_v.}
\end{eqnarray}

\subsection{Invariant schemes}

Before computing a set of fundamental discrete invariants
let us mention the fact that the equation 
\begin{equation}
T_+=0\label{weakly invariant burgers}
\end{equation}
is again weakly invariant under the group corresponding to
\eref{burgers algebra}.  The set of elementary invariants involving
the same discrete points as for the heat equation \eref{14 
dimensional space} is
\begin{eqnarray}
\eqalign{
I_1&=\frac{h_+}{h_-},\qquad\qquad%
I_2=\frac{\hat{h}_+}{\hat{h}_-},\qquad\qquad%
I_3=\frac{h_+\hat{h}_+}{\tau},\\
I_4&=h_+ h_-(v^+_x -v_x^-),\qquad%
I_5=\hat{h}_+ \hat{h}_-(\hat{v}^+_x -\hat{v}_x^-),\\
I_6&=h_+\Big( \frac{\sigma}{\tau} - v \Big),\qquad\quad\;\,%
I_7=\hat{h}_+\Big( \frac{\sigma}{\tau} - \hat{v} \Big),\\
I_8&=h_+^2 \Big( v^+_x + \frac{1}{\tau} \Big),\qquad\;\;\;
I_9=\hat{h}_+^2 \Big( \hat{v}^+_x - \frac{1}{\tau} \Big),}
\label{burgers invariants}
\end{eqnarray} 
where 
\begin{equation}
\fl v_x^+=\frac{v_+-v}{h_+},\qquad v_x^-=\frac{v-v_-}{h_-},\qquad
\hat{v}_x^+=\frac{\hat{v}_+-\hat{v}}{\hat{h}_+},\qquad
\hat{v}_x^-=\frac{\hat{v}-\hat{v}_-}{\hat{h}_-}.
\end{equation}

From the set of invariants \eref{burgers invariants} and
the weakly invariant equation \eref{weakly invariant burgers}
we derive invariant explicit and implicit schemes.
The explicit scheme is obtained by putting
\begin{equation}
(2I_6-I_7)I_3-I_8I_6=I_4,\qquad
T_+=0,\qquad
I_1=1,
\end{equation}
or in terms of the discrete variables
\numparts
\begin{eqnarray}
\bigg( 2\Big( \frac{\sigma}{\tau}-v \Big) +
\frac{\hat{h}}{h}\Big( \hat{v}-\frac{\sigma}{\tau} \Big) \bigg)
\frac{\hat{h}}{h\tau}+\Big( v_x^++\frac{1}{\tau} \Big)
\Big( v-\frac{\sigma}{\tau} \Big) =\frac{v_x^+-v_x^-}{h},
\label{burgers explicit u}\\
T_+=0,\label{burgers explicit t}\\
h_+=h_-\equiv h.\label{burgers explicit x}
\end{eqnarray}
\endnumparts

The implicit scheme is obtained by putting
\begin{equation}
(I_6-2I_7)I_3-I_9I_7=I_5,\qquad
T_+=0,\qquad
I_2=1,
\end{equation}
i.e.
\numparts
\begin{eqnarray}
\bigg( \frac{h}{\hat{h}} \Big( \frac{\sigma}{\tau}-v \Big) -
2 \Big( \frac{\sigma}{\tau}-\hat{v} \Big)\bigg) \frac{h}{\hat{h}\tau}-
\Big( \frac{\sigma}{\tau}-\hat{v} \Big) \Big( \hat{v}_x^+-\frac{1}{\tau}
\Big) =\frac{\hat{v}_x^+-\hat{v}_x^-}{h},\label{burgers implicit u}\\
T_+=0,\\
\hat{h}_+=\hat{h}_-\equiv \hat{h}.\label{burgers implicit x}
\end{eqnarray}
\endnumparts

As for the heat equation, the schemes generated involve
$\hat{h}$ and $h$.  Furthermore, the computation of the continuous
limit of the invariant schemes (59) and 
(61) gives the same condition on
the ratio $\sigma/\tau$ as for the heat equation, namely 
\eref{limit condition}.  To see this, we compute the 
continuous limit of the explicit scheme (59).  
The calculation is similar for the implicit scheme (61).
As for the heat equation, we
suppose that the steps in $t$ and $x$ are given by
\eref{parameter tau and h} and \eref{parameter sigma},
where we immediately assume that $k,l\geq 1$ 
in \eref{parameter sigma} .
First of all, it is clear that in the limit the equations specifying the lattice
\eref{burgers explicit t} and \eref{burgers explicit x} go to the identity
$0=0$.   The development of \eref{burgers explicit u} 
in a Taylor series in terms of the infinitesimal parameters $\epsilon$ and
$\delta$ gives

\begin{eqnarray}
\eqalign{
\fl\frac{\sigma}{\tau^2}
+\frac{\sigma}{\tau}\left( \frac{\sigma_+ -\sigma}{h\tau}\right)
-\frac{v}{\tau}+\frac{\sigma}{\tau}v_x
+v_t
-v\left(\frac{\sigma_+ -\sigma}{h\tau}\right)
+v\left(\frac{\sigma_+ -\sigma}{h\tau}\right)\\
-\frac{\sigma}{\tau}\left( \frac{\sigma_+ -\sigma}{h\tau}\right)
+vv_x
-\frac{\sigma}{\tau}v_x
+\frac{v}{\tau}-\frac{\sigma}{\tau^2}=
v_{xx}+\Or(\epsilon^{2k-1},\epsilon^{2l-1},\delta).}
\label{taylor series burgers}
\end{eqnarray} 
In  \eref{taylor series burgers} we have explicitely written all
the terms that do not converge to zero in the limit, but after canceling
the superfluous terms and making $\epsilon$ and $\delta$ go to zero
we retrieve Burgers' equation \eref{burgers equation}. 

In the particular case where the steps in $t$ are constant and
$\sigma$ is zero, the invariant discretization reduces to standard one,
namely
\begin{eqnarray*}
\frac{\hat{v}-v}{\tau}+vv_x^+=\frac{v_x^+-v_x^-}{h},\\
t=\tau m +t_0,\qquad x=h x + x_0,
\end{eqnarray*}
for the explicit scheme and
\begin{eqnarray*}
\frac{\hat{v}-v}{\tau}+\hat{v}\hat{v}_x^+=
\frac{\hat{v}_x^+-\hat{v}_x^-}{h},\\
t=\tau m +t_0,\qquad x=h x + x_0,
\end{eqnarray*}
for the implicit one.

The invariant schemes just generated are closely related to those 
obtained by Dorodnitsyn and Kozlov \cite{DK-1997}.  The invariant
schemes (59) and (61)
are not uniquely defined, the discrete approximations of the
Burgers' equation \eref{burgers explicit u}
and \eref{burgers implicit u} can be defined on other lattices.
For the explicit scheme we can replace  \eref{burgers explicit x}
by the invariant equation
\begin{equation}
I_6=0.
\end{equation}
On this new lattice,  \eref{burgers explicit u} becomes
\begin{equation}
-I_7I_3=I_4.
\end{equation}
In terms of the variables $(x,t,v)$ we get the invariant explicit scheme
\begin{eqnarray}
\eqalign{
\frac{\hat{v}-v}{\tau}=\frac{h_-}{\hat{h}_+^2}(v_x^+-v_x^-),\\
T_+=0,\\
\sigma=\tau v.}
\end{eqnarray}
For the implicit symmetry-preserving scheme we can replace
\eref{burgers implicit x} by $I_7=0$ to obtain a scheme similar 
to those in \cite{DK-1997}.   In the case of the Burgers' equation
the invariant schemes in \cite{DK-1997} can be seen as 
particular cases of the schemes we obtained.
The major difference between the
invariant schemes in \cite{DK-1997}
and (59) and (61) is that the steps in $x$ do 
not to have to be uniform in general.  

\subsection{Exact solutions}

Similarly as for the heat equation, it is clear that
the constant solution
\numparts
\begin{equation}
v(x,t)=v_0,\label{cste solution burgers}
\end{equation}
defined on the orthogonal lattice
\begin{equation}
t_{m,n}=\gamma(m),\qquad x_{m,n}=hn+x_0,\label{cste lattice burgers}
\end{equation}
\endnumparts
is an exact solution of the invariant schemes (59) and
(61) ($v_0$, $h$ and $x_0$ are constants).  
By choosing $\gamma(m)=\tau m +t_0$ we get a standard rectangular lattice.
Since the symmetry group of the Burgers' equation
is not very rich, just one new solution can be obtained from the
constant one.  By applying successively the symmetry
transformations $\exp[\epsilon\V_5]$, $\exp[v_0/\epsilon\V_1]$
and $\exp[1/\epsilon\V_2]$ to \eref{cste solution burgers} and
\eref{cste lattice burgers} we obtain
\numparts
\begin{equation}
\widetilde{v}=\widetilde{x}/\widetilde{t}\label{galilee solution}
\end{equation}
definied on the lattice
\begin{equation}
\widetilde{x}_{m,n}=\frac{\epsilon x_{m,n}+ (1-\epsilon t_{m,n})v_0}
{\epsilon(1-\epsilon t_{m,n})},
\qquad \widetilde{t}_{m,n}= \frac{1}{\epsilon (1-\epsilon t_{m,n})}
\end{equation}
\endnumparts
If we choose $v_0=0$, take $\epsilon=1$ 
and pose $\gamma(m)=1-1/(\tau m +t)$ in \eref{cste lattice burgers},
we conclude that the solution \eref{galilee solution} is an
exact solution of the invariant schemes on the lattice
\begin{equation}
t_{m,n}=\tau m +t_0,\qquad x_{m,n}=(hn+x_0)(\tau m +t_0).
\end{equation}  
The same lattice over which the fundamental solution of the heat equation
is exact.

\subsection{Burgers' equation in potential form}

The invariant discretization of the Burgers' equation in
potential form 
\begin{equation} 
w_t+\frac{(w_x)^2}{2}=w_{xx}\label{burgers potential}
\end{equation}
is readily obtained from that of the linear heat equation.  Indeed, since the PDE \eref{burgers
potential} is related to the heat equation \eref{heat equation}
by the point transformation
\begin{equation}
u=\exp\left[-\frac{w}{2}\right],
\label{transformation potential heat}
\end{equation}
the results of Section \ref{heat section} can directly
be used to generate invariant schemes and exact solutions via
the transformation \eref{transformation potential heat}.
For instance, the symmetry algebra of 
\eref{burgers potential} is spanned by 
\numparts
\begin{eqnarray}
\eqalign{
\V_1=\partial_x,\qquad
\V_2=\partial_t,\qquad
\V_3=t \partial_x + x \partial_w,\qquad
\V_4=x \partial_x + 2t \partial_t,\\
\V_5=tx \partial_x + t^2 \partial_t + (\frac{x^2}{2}+t)
\partial_w,\qquad
\V_6=\partial_w,}
\label{finite algebra burgers potential}
\end{eqnarray}
\begin{equation}
\V_\alpha=\alpha(x,t)\exp\left[\frac{w}{2}\right] \partial_w,
\label{infinite burgers potential}
\end{equation}
\endnumparts
where $\alpha$ is a solution of the heat equation, $\alpha_t=\alpha_{xx}$.

\subsubsection{Invariant schemes}

Since the transformation \eref{transformation potential heat}
does not affect the independant variables, it is clear that the
equation \eref{heat invariant manifold} is still a weakly invariant
equation.  We can directly use the set of elementary
invariants \eref{heat invariants} and the transformation
\eref{transformation potential heat} to obtain a set of
elementary invariants on a flat time layer scheme \cite{V-2005-2}
\begin{eqnarray}
\fl \eqalign{
I_1=\frac{h_+}{h_-},\qquad
I_2=\frac{\hat{h}_+}{\hat{h}_-},\qquad
I_3=\frac{h_+\hat{h}_+}{\tau}, \qquad
I_4=\frac{\sqrt{\tau}}{h_+}\exp\left[
-\frac{\hat{w}-w}{2}+\frac{\sigma^2}{4\tau}\right]\\
I_5=\exp\left[-\frac{w_+-w}{2}+
\frac{h_+}{4\tau}(2\sigma-h_+)\right], \qquad
I_6=\exp\left[\frac{w-w_-}{2}
-\frac{h_-}{4\tau}(2\sigma+h_-)\right],\\
I_7=\exp\left[-\frac{\hat{w}_+-\hat{w}}{2}+ 
\frac{\hat{h}_+}{4\tau}(2\sigma+\hat{h}_+)\right],\qquad
I_8=\exp\left[\frac{\hat{w}-\hat{w}_-}{2}
-\frac{\hat{h}_-}{4\tau}
(2\sigma-\hat{h}_+)\right].}
\end{eqnarray}
By using the same invariant expressions 
\eref{invariant explicit heat} and
\eref{invariant implicit heat}
we obtain invariant schemes for \eref{burgers potential}.
An explicit scheme is
\begin{eqnarray}
\eqalign{
\fl\frac{1}{\tau}\bigg( \sqrt{\frac{\hat{h}}{h}}
\exp\left[-\frac{\hat{w}}{2}+\frac{\sigma^2}{4\tau}\right]-
\exp\left[-\frac{w}{2}\right] \bigg) =
\frac{1}{\hat{h}h} \bigg(
\exp\left[ -\frac{w_+}{2}+\frac{h}{4\tau}
(2\sigma -h+\hat{h})\right] \\
-2\exp\left[-\frac{w}{2}\right]
+\exp\left[\frac{w_-}{2}-\frac{h}{4\tau}
(2\sigma +h-\hat{h})\right] \bigg),\\
T_+=0,\\
h_+=h_-\equiv h.}\label{explicit burgers potential}
\end{eqnarray}
An implicit one is given by
\begin{eqnarray}
\eqalign{
\fl \frac{1}{\tau}\bigg( \exp\left[-\frac{\hat{w}}{2}\right]-
\sqrt{\frac{h}{\hat{h}}}
\exp\left[-\frac{w}{2}-\frac{\sigma^2}{4\tau}\right]\bigg) =
\frac{1}{\hat{h}h}\bigg(
\exp\left[-\frac{w_+}{2}+\frac{\hat{h}}{4\tau}(2\sigma-
h+\hat{h})\right]\\
-2\exp\left[-\frac{\hat{w}}{2}\right]
+\exp\left[-\frac{\hat{w}_-}{2}
-\frac{h}{4\tau}(2\sigma +h
-\hat{h})\right]\bigg) ,\\
T_+=0,\\
\hat{h}_+=\hat{h}_-\equiv \hat{h}.}\label{implicit burgers potential}
\end{eqnarray}

The invariant schemes \eref{explicit burgers potential} and
\eref{implicit burgers potential} converge to
the differential equation \eref{burgers potential}, 
if the condition \eref{limit condition} on the lattice is
satisfied.  In contrast to the invariant schemes for the
heat equation (32) and 
\eref{implicit heat}, which give the standard discretization
when $\sigma=0$; the invariant schemes
\eref{explicit burgers potential} and 
\eref{implicit burgers potential} give
unusual schemes on a rectangular lattice.  For example,
when $\sigma=0$ the explicit scheme \eref{explicit burgers potential}
becomes on a rectangular lattice
\begin{eqnarray}
\eqalign{
\frac{\exp\left[-\frac{\hat{w}}{2}\right]-
\exp\left[-\frac{w}{2}\right]}{\tau}=
\frac{\exp\left[-\frac{w_+}{2}\right]-
2\exp\left[-\frac{w}{2}\right]+\exp\left[-\frac{w_-}{2}\right]}
{h^2},\\
t=\tau m+t_0,\qquad x=h n+x_0,}
\end{eqnarray}
which is quit different from the standard discretization
\begin{displaymath}
\frac{\hat{w}-w}{\tau}+\frac{1}{2}\left(\frac{w_+-w}{h}\right)^2=
\frac{w_+-2w+w_-}{h^2}.
\end{displaymath}

Finally, let us mention that the exact solutions obtained in Section
\ref{exact solution heat} are mapped by \eref{transformation potential heat}
to exact solutions of the invariant schemes \eref{explicit burgers potential}
and \eref{implicit burgers potential}.  Since the transformation does not
affect the independant variables the solutions obtained via 
\eref{transformation potential heat} are exact on the same lattice.

\section{Korteweg-de Vries equation}\label{kdv section}

As a last example, we consider the Korteweg-de Vries equation
\begin{equation}
u_t=uu_x+u_{xxx}.\label{kdv}
\end{equation}
This equation has a four dimensional symmetry algebra
spanned by
\begin{equation}
\fl\V_1=\partial_x,\qquad
\V_2=\partial_t,\qquad
\V_3=t \partial_x - \partial_u,\qquad
\V_4=x \partial_x + 3t \partial_t - 2u\partial_u.
\label{kdv algebra}
\end{equation}

\subsection{Invariant schemes}

Again, the equation \eref{heat invariant manifold} is
weakly invariant and we can
discretize \eref{kdv} on a lattice with 
flat time layers. This time we make use of all
points shown in figure \ref{figure lattice}
in order to approximate correctly the third order derivative
in $x$.  The usual computation gives
the invariants\\
\begin{eqnarray}
I_1=\frac{h_+}{h_-},&
I_2=\frac{h_{++}}{h_+},&
I_3=\frac{h_-}{h_{--}},\nonumber\\
I_4=\frac{\hat{h}_+}{\hat{h}_-},&
I_5=\frac{\hat{h}_{++}}{\hat{h}_+},&
I_6=\frac{\hat{h}_-}{\hat{h}_{--}},\nonumber\\
I_7=\frac{h_+}{\hat{h}_+},&
I_8=\frac{h_+^3}{\tau},&
I_9=\frac{\sigma+\tau u}{h_+},\\
I_{10}=\tau u_x^{--},\qquad&
I_{11}=\tau u_x^-,&
I_{12}=\tau u_x^+,\nonumber\\
I_{13}=\tau u_x^{++},&
I_{14}=h_+^2(\hat{u}-u),\qquad&
I_{15}=\tau \hat{u}_x^{--},\nonumber\\
I_{16}=\tau \hat{u}_x^-,&
I_{17}=\tau \hat{u}_x^+,&
I_{18}=\tau \hat{u}_x^{++},  \nonumber
\end{eqnarray}
with
\begin{equation}
\fl
u_x^{++}=\frac{u_{++}-u_+}{h_{++}},\quad
u_x^{--}=\frac{u_--u_{--}}{h_{--}},\quad
\hat{u}_x^{++}=\frac{\hat{u}_{++}-\hat{u}_+}{\hat{h}_{++}},\quad
\hat{u}_x^{--}=\frac{\hat{u}_--\hat{u}_{--}}{\hat{h}_{--}}.
\end{equation}

An invariant explicit scheme approximating \eref{kdv} is obtained
by putting
\begin{equation}
\fl I_{14}=I_9I_8\frac{I_{12}-I_{11}}{2}+\frac{1}{2}
(I_{13}-I_{12}-I_{11}+I_{10}),\qquad
T_+=0,\qquad
I_1=1.
\end{equation}
In terms of the original variables we have
\numparts
\begin{eqnarray}
\frac{\hat{u}-u}{\tau}=u\frac{u_+-u_-}{2h}+
\frac{u_{++}-2u_++2u_--u_{--}}{2h^3}+\frac{\sigma}{\tau}
\frac{u_+-u_-}{2h},\label{kdv explicit u}\\
T_+=0,\\
h_+=h_-\equiv h.\label{kdv explicit x}
\end{eqnarray}
\endnumparts

Furthermore, an implicit scheme is obtained 
by considering the combination of invariants 
\begin{equation}
\fl I_{14}=(I_9+I_8^{-1}I_{14})I_8\frac{I_{17}-I_{16}}{2}+
\frac{1}{2}(I_{18}-I_{17}-I_{16}+I_{15})I_7^2,\quad
T_+=0,\quad
I_4=1,
\end{equation}
i.e.
\numparts
\begin{eqnarray}
\frac{\hat{u}-u}{\tau}=\hat{u}\frac{\hat{u}_+-\hat{u}_-}
{2\hat{h}}+
\frac{\hat{u}_{++}-2\hat{u}_++2\hat{u}_--\hat{u}_{--}}{2\hat{h}^3}+
\frac{\sigma}{\tau}
\frac{\hat{u}_+-\hat{u}_-}{2\hat{h}},
\label{kdv implicit u}\\
T_+=0,\\
\hat{h}_+=\hat{h}_-\equiv \hat{h}.\label{kdv implicit x}
\end{eqnarray}
\endnumparts

The invariant schemes include 
the standard discretizations.  Indeed, if $\sigma$ equals 
zero then the last term on the right hand side of \eref{kdv explicit u} 
and \eref{kdv implicit u} vanishes, 
which is just the usual discretization of
the Korteweg-de Vries equation.  In order for the equations 
\eref{kdv explicit u} and \eref{kdv implicit u} to converge
to \eref{kdv} the condition
\begin{displaymath}
\lim_{\tau\to 0} \frac{\sigma}{\tau}<\infty
\end{displaymath}
must be imposed.

As for the Burgers' equation, by considering the invariant schemes (81) and (83) on a lattice depending on the
solution we recover similar schemes as those obtained in \cite{D-1994}.
For the explicit scheme (81), if we replace \eref{kdv explicit x} by 
\begin{equation}
I_9=0,
\end{equation}
the equation \eref{kdv explicit u} then becomes
\begin{equation}
I_{14}=\frac{1}{2}(I_{13}-I_{12}-I_{11}+I_{10}).
\end{equation}
In term of the variables $(x,t,u)$ we have
\begin{eqnarray}
\eqalign{
\frac{\hat{u}-u}{\tau}=\frac{(u_x^{++}-u_x^+)-(u_x^--u_x^{--})}{h_+^2},\\
T_+=0,\\
\sigma=-\tau u.}
\end{eqnarray}
For the implicit scheme (83) we just have to replace
\eref{kdv implicit x} by $I_9+I_{14}I_8^{-1}=0$
to get the invariant implicit scheme
\begin{eqnarray}
\eqalign{
\frac{\hat{u}-u}{\tau}=\frac{(\hat{u}_{x}^{++}-\hat{u}_x^+)
-(\hat{u}_x^--\hat{u}_x^{--})}{\hat{h}_+^2}\\
T_+=0,\\
\sigma=-\tau\hat{u}.}
\end{eqnarray}

\subsection{Exact solutions}

As with all previously considered invariant schemes, the 
constant solution defined on an orthogonal mesh is
an exact solution.  Another exact solution of the invariant
schemes corresponds to the solution invariant under
the Galilei transformation generated by the vector
field $\V_3$.  This exact solution is obtained in
a non-obvious fashion.  The idea is the following one.
Given the partial differential equation that we approximate, 
we can perform a symmetry reduction to obtain an exact solution
$u(x,t)$.
Given this solution, we can insert it in the discrete equation
approximating the original differential equation, which
will give an equation relating the independant variables.
If the equation obtained can be solved using the existing
liberty in the lattice, then we have generated an exact
solution of the P$\Delta$S.

In the continuous case solution invariant
under the infinitesimal symmetry generator $\V_3$ is
\begin{equation}
u=-\frac{x}{t}.\label{invariant kdv u}
\end{equation}
By replacing this solution in \eref{kdv explicit u},
we get
\begin{equation}
x=\frac{\sigma t}{\tau}.\label{invariant kdv lattice}
\end{equation}
If we take the evolution in time to be
\begin{equation}
t_{m,n}=\tau m+t_0,\label{invariant kdv lattice t}
\end{equation}
the equation \eref{invariant kdv lattice} 
can be solved for $x$ and gives
\begin{equation}
x_{m,n}=(hn+x_0)(\tau m+t_0).\label{invariant kdv lattice x}
\end{equation}
Hence, we conclude that \eref{invariant kdv u}
is an exact solution of the explicit scheme (81)
on the lattice defined by the equations 
\eref{invariant kdv lattice t} and \eref{invariant kdv lattice x}.

\section{CONCLUSIONS}\label{conclusions section} 

The discretization procedure presented in Section \ref{section discretization}
and applied to specific examples in Section \ref{heat section}, \ref{burgers section}
and \ref{kdv section} is a ``minimal'' one.  By that we mean that the 
PDE \eref{pde} is replaced by a P$\Delta$S \eref{pfde} involving
only $N=3$ difference equations.  The P$\Delta$S is by construction
invariant under the symmetry group of the original PDE.  The general
solution of the P$\Delta$S \eref{pfde} will depend on several arbitrary
functions of one (discrete) variable.  A specific solution $u(x,t)$ and a 
specific lattice on which the solution is valid is obtained once these arbitrary functions
are specified.  

The freedom inherent in these arbitrary functions can be used in different ways.  For 
instance, we can require that certain physically important solutions of the 
PDE should also be exact solutions of the P$\Delta$S.  This was done in Section
\ref{heat section} with the fundamental solution of the heat equation.

An alternative possibility is to use the invariants of the symmetry group $G$ to
add further invariant equations to the ``minimal'' P$\Delta$S \eref{pfde}.
This creates an overdetermined system of difference equations, restricting
the freedom in the solutions.  Since the system of $N$ equations
with $N\geq 4$ is overdetermined, their compatibility must be assured.
The idea of imposing an overdetermined system of invariant
difference equations has already been explored earlier
\cite{D-1991,DK-2003,LTW-2001} with the aim of further restricting
the form of the lattice.

All PDEs studied in this article are evolution equations of the
form
\begin{equation}
u_t=f(x,t,u,u_x,u_{xx},u_{xxx})
\end{equation}
and we made use of the fact that their symmetry groups always allowed
horizontal time layers.

Other types of equations, requiring different lattices, will be
studied in the future.  The extension to more than one dependent
variable and more than two independant ones is obvious.

This research program, ``continuous symmetries of discrete equations'' has 
both physical and mathematical aspects that must be further pursued.
One conclusion is that different discrete physical systems require
different approaches.  It was shown elsewhere \cite{LTW-,LTW-2004}
that symmetries of linear theories can be studied in terms of 
commuting difference operators on fixed lattices.  The results
of this and related papers \cite{BDK-1997,BD-2001,D-1991,D-1994,D-2001,DK-1997,DK-2003,DKW-2000,DKW-2004,DW-2000,LTW-2000,LTW-2001,LW-1991,RW-2004} indicate that at least for nonlinear
discrete phenomena, the lattice should be considered as a dynamical
one evolving together with the solution and described by an invariant
system of difference equations.

From the mathematical point of view we see as the greatest challenge
the development of symmetry adapted numerical schemes that give
better results than standard numerical methods.

\ack

We thank Anne Bourlioux, Vladimir Dorodnitsyn and Decio Levi for
many interesting and helpful discussions.  The research of P.W.
was partially supported by NSERC of Canada.  F.V. thanks NSERC for
a Canada graduate scholarship.

\section*{References}



\begin{thebibliography}{99}

\bibitem{BDK-1997}
Bakirova M I, Dorodnitsyn V A and Kozlov R V 1997 Symmetry preserving
difference schemes for some heat transfer equations {\it J. Phys A. Math. Gen}
{\bf 30} 8139--55 

\bibitem{BD-2001}
Budd C J and Dorodnitsyn V A 2001 Symmetry adapted moving mesh
schemes for the nonlinear Schr\"odinger equation, J. Phys. A. Math. Gen.
{\bf 34} 10387--400 

\bibitem{D-1991}
Dorodnitsyn V A 1991 Transformation groups in a space of difference variables
{\it J. Sov. Math.} {\bf 55} 1490--517

\bibitem{D-1994}
Dorodnitsyn V A 1994 Invariant discrete model for the Korteweg-de Vries
equation {\it Preprint CRM-2187, Universit\'e de Montr\'eal}

\bibitem{D-2001}
Dorodnitsyn V A 2001 Group properties of difference equations (Fizmatlit,
Moscow) (in Russian)

\bibitem{DK-1997}
Dorodnitsyn V A and Kozlov R 1997 The whole set of symmetry preserving
discrete versions of a heat transfer equation with a source {\it SYNODE,
preprint Numerics No. 4}

\bibitem{DK-2003}
Dorodnitsyn V A and Kozlov R 2003 Heat transfer with a source:
the complete set of invariant difference schemes
{\it J. Nonlin. Math. Phys} {\bf 10} 16--50 

\bibitem{DKW-2000}
Dorodnitsyn V A,  Kozlov R and Winternitz P 2000 Lie group
classification of second order ordinary difference
equations {\it J. Math. Phys.} {\bf 41} 480--504

\bibitem{DKW-2004}
Dorodnitsyn V A, Kozlov R and Winternitz P 2004 Continuous
symmetries of Lagrangians and exact solutions of discrete
equations {\it J. Math. Phys.} {\bf 45} 336--59

\bibitem{DW-2000}
Dorodnitsyn V A and Winternitz P 2000 Lie point symmetry
preserving discretizations for variable coefficient
Korteweg-de Vries equations {\it Nonlinear Dynamics} {\bf 22}
49--59

\bibitem{FNNV-1996}
Floreanini R, Negro J, Nieto L M and Vinet L 1996 Symmetries of the heat
equation on a lattice {\it Lett. Math. Phys.} {\bf 36} 351--55 

\bibitem{LTW-}
Levi D, Tempesta P and Winternitz P 2004 Lorentz and Galilei invariance
on Lattices {\it Phys. Rev. D} {\bf 69} 105011 pp. 1--6

\bibitem{LTW-2004}
Levi D, Tempesta P and Winternitz P 2004 Umbral calculus, difference
equations and the discrete Schr\"odinger equation {\it J. Math. Phys.}
{\bf 45} 4077--105

\bibitem{LTW-2000}
Levi D, Tremblay S and Winternitz P 2000 Lie point symmetries of
difference equations and lattices {\it J. Phys. A Math. Gen.} {\bf 33}
8507--24 

\bibitem{LTW-2001}
Levi D, Tremblay S and Winternitz P 2001 Lie symmetries of multidimensional
equations and lattices {\it J. Phys. A Math. Gen.} {\bf 34} 9507--24 

\bibitem{LVW-1997}
Levi D, Vinet L and Winternitz P 1997 Lie group formalism for difference
equations {\it J. Phys. A Math. Gen.} {\bf 30} 633--49 

\bibitem{LW-1991}
Levi D and Winternitz P 1991 Continuous symmetries of discrete equations
{\it Phys. Lett. A} {\bf 152} 335--38 

\bibitem{LW-2005}
Levi D and Winternitz P 2004 Continuous symmetries of difference
equations {\it Preprint nlin.SI/0502004}

\bibitem{M-1987}
Maeda S 1987 The similarity method for difference equations {\it IMA J. Appl. Math.}
{\bf 38} 129--34 

\bibitem{QCS-1992}
Quispel G R W, Capel H W and Sahdevan R 1992 Continuous
symmetries of difference equations:  the Kac-van Morerbeke equation
and the Painlev\'e reduction {\it Phys. Lett. A} {\bf 170} 379--83

\bibitem{RW-2004}
Rodriguez M A and Winternitz, P 2004 Lie symmetries and exact solutions of
first-order difference schemes {\it J. Phys. A Math. Gen.} {\bf 37} 6129--42 

\bibitem{V-2005-1}
Valiquette F 2005 Discretizations preserving all Lie point symmetries of the 
Korteweg-de Vries equation {\it Preprint math-ph/0507033}

\bibitem{V-2005-2}
Valiquette F 2005 Point transformation in invariant difference schemes
{\it Preprint math-ph/0507041}

\bibitem{W-2004}
Winternitz P 2004 Symmetries of discrete systems, in {\it Discrete Integrable
Systems, Lecture Notes in Physics 644}, eds. Grammaticos B. 
Kossmann-Schwarzbach Y and Tamizhmani T (Springer, Berlin) 185--243

\bibitem{O-1993}
Olver P J 1993 {\it Applications of Lie Groups to Differential 
Equations} (Springer, New York) 

\bibitem{O-1995}
Olver P J 1995 {it Equivalence, Invariants, and 
Symmetries} (Cambridge University Press, Cambridge) 

\bibitem{FO-1998}
Fels M and Olver P J 1998 Moving coframes $\textrm{I}$.  A practical
algorithm {\it Acta Applic. Math.} {\bf 51} 161--213

\bibitem{O-2003}
Olver P J 2003 Moving frames {\it J. Symb. Comp.}  {\bf 36} 501--12


\end{thebibliography}
\end{document}